\documentclass{llncs}

\usepackage{amsmath}
\usepackage{amssymb}
\usepackage{graphicx}
\usepackage[hyphens]{url}
\usepackage{bm}
\usepackage{textcomp}
\newcommand{\V}{\mathbf{v}}

\begin{document}

\mainmatter
%================================================

%==== FILL IN ====================================
\title{Non-linear Real Arithmetic Benchmarks derived from Automated Reasoning in Economics}  % Full title
\titlerunning{NRA Benchmarks derived from Automated Reasoning in Economics} % Short title
\author{Casey B. Mulligan\inst{1} \and  Russell Bradford \inst{2} \and James H. Davenport \inst{2} \and \\ Matthew England \inst{3} \and Zak Tonks \inst{2}}
\authorrunning{Mulligan-Bradford-Davenport-England-Tonks}
\institute{
University of Chicago, USA \\%$-$
\email{c-mulligan@uchicago.edu}\\ 
%\texttt{url}
\and
University of Bath, UK \\%$-$
\email{\{R.J.Bradford, J.H.Davenport, Z.P.Tonks\}@bath.ac.uk}\\ 
%\texttt{url}
\and
Coventry University, UK \\%$-$
\email{Matthew.England@coventry.ac.uk}\\ 
%\texttt{url}
}
\maketitle

\begin{abstract}
We consider problems originating in economics that may be solved automatically using mathematical software.  We present and make freely available a new benchmark set of such problems.
The problems have been shown to fall within the framework of non-linear real arithmetic, and so are in theory soluble via Quantifier Elimination (QE) technology as usually implemented in computer algebra systems.  Further, they all can be phrased in prenex normal form with only existential quantifiers and so are also admissible to those Satisfiability Module Theory (SMT) solvers that support the \verb+QF_NRA+ logic.
There is a great body of work considering QE and SMT application in science and engineering, but we demonstrate here that there is potential for this technology also in the social sciences.
\end{abstract}

%------------------------------------------------------------------------------

\section{Introduction}
\label{SEC:Intro}

\subsection{Economic reasoning}
\label{SUBSEC:Framework}

A general task in economic reasoning is to determine whether, with variables $\V=(v_1,\ldots,v_n)$, the hypotheses $H(\V)$ follow from the assumptions $A(\V)$, i.e. is it the case that
\[
\forall\V \, . \, A(\V) \Rightarrow H(\V)?
\]
Ideally the answer would be \verb+True+ or \verb+False+, and of course logically it is.  But the application being modelled, like real life, can require a more nuanced analysis.  It may be that for most $\V$ the theorem holds but there are some special cases that should be ruled out (additions to the assumptions).  Such knowledge is valuable to the economist.  Another possibility is that the particular set of assumptions chosen are contradictory\footnote{A lengthy set of assumptions is many times required to conclude the hypothesis of interest so the possibility of contradictory assumptions is real.}, i.e. $A(\V)$ itself is \verb+False+.  As all students of an introductory logic class know, this would technically make the implication \verb+True+, but for the purposes of economics research it is important to identify this separately!  

We categorise the situation into four possibilities that are of interest to an economist (Table \ref{tab:main}) via the outcome of a pair of fully existentially quantified statements which check the existence of both an example ($\exists \V[A \land H]$) and counterexample ($\exists \V[A \land \lnot H]$) of the theorem.  So we see that every economics theorem generates a pair of SAT problems, in practice actually a trio since we would first perform the cheaper check of the compatibility of the assumptions ($\exists \V[A]$).

\begin{table}[t]
\caption{Table of possible outcomes from a potential theorem $\forall \, . \, \V A \Rightarrow H$ \label{tab:main} }
\centering
\begin{tabular}{r|cc}
& $\lnot \exists \V[A \land \lnot H]$ & $\exists \V[A \land \lnot H]$ \\[0.1cm] \hline 
$\exists \V[A \land H]$      \,& \verb+True+          & \verb+Mixed+ \\
$\lnot \exists \V[A \land H]$ \, & \, Contradictory Assumptions & \verb+False+
\end{tabular} 
\end{table}

Such a categorisation is valuable to an economist.  They may gain a proof of their theorem, but if not they still have important information to guide their research: 
knowledge that their assumptions contradict; or information about 
$
\{\V : A(\V)\Rightarrow H(\V)\}.
$

Also of great value to an economist are the options for exploration that such technology would provide.  An economist could vary the question by strengthening the assumptions that led to a \verb+Mixed+ result in search of a \verb+True+ theorem.  However, of equal interest would be to weaken the assumption that generated a \verb+True+ result for the purpose of identifying a theorem that can be applied more widely.  
Such weakening or strengthening is implemented by quantifying more or less of the variables in $\V$.  For example, we might partition $\V$ as $\V_1,\V_2$ and ask for $\{\V_1:\forall \V_2 \, . \, A(\V_1,\V_2)\Rightarrow H(\V_1,\V_2)\}$. The result in these cases would be a formula in the free variables that weakens or strengthens the assumptions as appropriate.  

\subsection{Technology to solve such problems automatically}
\label{SUBSEC:Technology}

Such problems all fall within the framework of \emph{Quantifier Elimination} (QE).  QE refers to the generation of an equivalent quantifier free formula from one that contains quantifiers.  SAT problems are thus a sub-case of QE when all variables are existentially quantified.  

QE is known to be possible over real closed fields (real QE) thanks to the seminal work of Tarski \cite{Tarski1948} and practical implementations followed the work of Collins on the Cylindrical Algebraic Decomposition (CAD) method \cite{Collins1975} and Weispfenning on Virtual Substitution \cite{Weispfenning1988}.  There are modern implementations of real QE in \textsc{Mathematica} \cite{Strzebonski2006}, \textsc{Redlog} \cite{DS97a}, \textsc{Maple} (\textsc{SyNRAC} \cite{IYA14} and the \textsc{RegularChains}  Library \cite{CM14c}) and \textsc{Qepcad-B} \cite{Brown2003b}.

The economics problems identified all fall within the framework of QE over the reals.  Further, the core economics problem of checking a theorem can be answered via fully existentially quantified QE problems, and so also soluble using the technology of Satisfiability Modulo Theory (SMT) Solvers; at least those that support the \verb+QF_NRA+ (Quantifier Free Non-Linear Real Arithmetic) logic such as \textsc{SMT-RAT} \cite{CLJA12}, \textsc{veriT} \cite{FOSV17}, \textsc{Yices2} \cite{JD17},  and \textsc{Z3} \cite{JdM12}.

\subsection{Case for novelty of the new benchmarks}
\label{SUBSEC:Novelty}

QE has found many applications within engineering and the life sciences.  
Recent examples include the derivation of optimal numerical schemes \cite{EH16}, 
artificial intelligence to pass university entrance exams \cite{WMTA16}, 
weight minimisation for truss design \cite{CC18}, 
and biological network analysis \cite{BDEEGGHKRSW17}.
However, applications in the social sciences are lacking (the nearest we can find is \cite{LW14}).  

Similarly, for SMT with non-linear reals, applications seem focussed on other areas of computer science.  The original \texttt{nlsat} benchmark set \cite{JdM12} was made up mostly of verification conditions from the \textsc{Keymaera} \cite{PQR09}, and theorems on polynomial bounds of special functions generated by the \textsc{MetiTarski} automatic theorem prover \cite{Paulson2012}.  This category of the SMT-LIB \cite{SMTLIB} has since been broadened with problems from physics, chemistry and the life sciences \cite{SW08}. However, we are not aware of any benchmarks from economics or the social sciences.

The reader may wonder why QE has not been used in economics previously.  On a few occasions when QE algorithms have been mentioned in economics they have been characterized as ``something that is do-able in principle, but not by any computer that you and I are ever likely to see'' \cite{Steinhorn2008}.  Such assertions were based on interpretations of theoretical computer science results rather than experience with actual software applied to an actual economic reasoning problem.  Simply put, the recent progress on QE/SMT technology is not (yet) well known in the social sciences.  

\subsection{Availability and further details}
\label{SUBSEC:Availability}

The benchmark set consists of 45 potential economic theorems.  Each theorem requires the three QE/SMT calls to check the compatibility of assumptions, the existence of an example, and the existence of a counterexample, so 135 problems in total.  In all cases the assumption and example checks are SAT, in fact often fully satisfied (any point can witness it) as they relate to a true theorem.  Correspondingly, the counterexample is usually UNSAT (for 42/45 theorems) and thus the more difficult problem from the SMT point of view.  

The benchmark problems are hosted on the Zenodo data repository at URL 
\url{https://doi.org/10.5281/zenodo.1226892} in both the SMT2 format and as input files suitable for \textsc{Redlog} and \textsc{Maple}. The SMT2 files have been accepted into the SMT-LIB \cite{SMTLIB} and will appear in the next release.

Available from \url{http://examples.economicreasoning.com/} are \textsc{Mathematica} notebooks\footnote{A \textsc{Mathematica} licence is needed to run them, but static pdf print outs are also available to download.} which contain commands to solve the examples in \textsc{Mathematica} and also further information on the economic background: meaning of variable names and economic implications of results etc.

\newpage

\subsection{Plan}

The paper proceeds as follows.  In Section \ref{SEC:Examples} we describe in detail some examples from economics, ranging from textbook examples common in education to questions arising from current research discussions.  Then in Section \ref{SEC:ExampleCollection} we offer some statistics on the logical and algebraic structure of these examples.  We then give some final thoughts on future and ongoing work in Section \ref{SEC:Final}.

%\newpage

\section{Examples of Economic Reasoning in Tarski's Algebra}
\label{SEC:Examples}

The fields of economics ranging from macroeconomics to industrial organization to labour economics to econometrics involve deducing conclusions from assumptions or observations.  Will a corporate tax cut cause workers to get paid more?  Will a wage regulation reduce income inequality?  Under what conditions will political candidates cater to the median voter?  We detail in this section a variety of such examples.

\subsection{Comparative static analysis}
\label{SUBSEC:Marshall}

We start with Alfred Marshall's \cite{Marshall1895} classic, and admittedly simple, analysis of the consequences of cost-reducing progress for activity in an industry.  Marshall  concluded that, for any supply-demand equilibrium in which the two curves have their usual slopes, a downward supply shift increases the equilibrium quantity $q$ and decreases the equilibrium price $p$.  

One way to express his reasoning in Tarski's framework is to look at the industry's local comparative statics: meaning a comparison of two different equilibrium states between supply and demand.  
With a downward supply shift represented as $da > 0$ (where $a$ is a reduction in costs) we have here:
\begin{align*}
&A \equiv D'(q)<0 \wedge S'(q)>0 \\
&\qquad \wedge \dfrac{d}{da}\big(S(q)-a\big)=\frac{dp}{da}
\wedge \frac{dp}{da}=\frac{d}{da}D(q)
\\
&H \equiv \frac{dq}{da}>0 \wedge \frac{dp}{da}<0
\end{align*}
where: 
\begin{itemize}
\item $D'(q)$ is the slope of the inverse demand curve in the neighborhood of the industry equilibrium;
\item $S'(q)$ is the slope of the inverse supply curve; 
\item $\dfrac{dq}{da}$ is the quantity impact of the cost reduction; and 
\item $\dfrac{dp}{da}$ is the price impact of the cost reduction.
\end{itemize}
Economically, the first atoms of $A$ are the usual slope conditions: that demand slopes down and supply slopes up.  The last two atoms of $A$ say that the cost change moves the industry from one equilibrium to another.  Marshall's hypothesis was that the result is a greater quantity traded at a lesser price.

Hence we set the ``variables'' $\V$ to be four real numbers $(v_1,v_2,v_3,v_4)$:
\[
\V = \left\{D'(q),S'(q),\dfrac{dq}{da},\dfrac{dp}{da}\right\}.
\]
Then, after applying the chain rule, $A$ and $H$ may be understood as Boolean combinations of polynomial equalities and inequalities:
\begin{align*}
&A \equiv v_1<0 \wedge v_2>0  \wedge v_3v_2-1=v_4 \wedge v_4=v_3v_1 ,
\\
&H \equiv v_3>0 \wedge v_4<0.
\end{align*}
Thus Marshall's reasoning fits in the Tarski framework and therefore is amenable to automatic solution.  Any of the tools mentioned in the introduction can easily evaluate the two existential problems for Table \ref{tab:main} and conclude easily that $\forall\V \, . \, A \Rightarrow H$ is \texttt{True}, confirming Marshall's conclusion.  

In fact, for this example it is straightforward to produce a short hand proof in a couple of lines\footnote{Subtract the final assumption from the penultimate one and then apply the sign conditions of the other assumptions to conclude $v_3>0$.  Then with this the last assumption provides the other hypothesis, $v_4<0$.} and so we did not include it in the benchmark set.  But it demonstrates the basic approach to placing economics reasoning into a form admissible for QE/SMT tools.  A similar approach is used for many of the larger examples forming the benchmark set, such as the next one.

\subsection{Scenario analysis}
\label{SUBSEC:Krugman}

Economics is replete with \emph{``what if?''} questions.  Such questions are logically and algebraically more complicated, and thereby susceptible to human error, because they require tracking various scenarios.  We consider now one such question originating from recent economic debate.
  
Writing at nytimes.com\footnote{\url{https://krugman.blogs.nytimes.com/2012/11/03/soup-kitchens-caused-the-great-depression/}}, Economics Nobel laureate Paul Krugman asserted that whenever taxes on labour supply are primarily responsible for a recession, then wages increase.  
Two scenarios are discussed here: what actually happens ($act$) when taxes ($t$) and demand forces ($a$) together create a recession, and what would have happened ($hyp$) if taxes on labour supply had been the only factor affecting the labour market.  

Expressed logically we have:
\begin{align*}
A &\equiv \bigg( \, \frac{\partial D(w,a)}{\partial w}<0 \wedge \frac{\partial S(w,t)}{\partial w}>0  
\\
&\qquad \wedge
	\frac{\partial D(w,a)}{\partial a}=1 \wedge
    \frac{\partial S(w,t)}{\partial t}=1 
\\
&\qquad \wedge
    \frac{d}{dact}\big(D(w,a)=q=S(w,t)\big)
\\
&\qquad  \wedge   
    \frac{d}{dhyp}\big(D(w,a)=q=S(w,t)\big)
\\
&\qquad \wedge
	\frac{d t}{dact}=\frac{d t}{dhyp}\wedge
	\frac{d a}{dhyp}=0
\\
&\qquad \wedge
	\frac{dq}{dhyp}<\frac{1}{2} \frac{dq}{dact}<0 \, \bigg)
\\
H &\equiv \frac{dw}{dact}>0.
\end{align*}
In economics terms, the first line of assumptions contains the usual slope restrictions on the supply and demand curves.  Because nothing is asserted about the units of $a$ or $t$, the next line just contains normalizations.  The third and fourth lines say that each scenario involves moving the labour market from one equilibrium (at the beginning of a recession) to another (at the end of a recession).  The fifth line defines the scenarios: both have the same tax change but only the $act$ scenario has a demand shift.  The final assumption / assertion is that a majority of the reduction in the quantity $q$ of labour was due to supply (that is, most of $\frac{dq}{dact}$ would have occurred without any shift in demand).  The hypothesis is that wages $w$ are higher at the end of the recession than they were at the beginning.

%\noindent 
Viewed as a Tarski formula this has twelve variables, 
\begin{align*}
\V = \bigg\{
&\frac{da}{dact},\frac{da}{dhyp},\frac{dt}{dact},\frac{dt}{dhyp}, 
\\
&\frac{dq}{dact},\frac{dq}{dhyp},\frac{dw}{dact},\frac{dw}{dhyp}, 
\\
&\frac{\partial D(w,a)}{\partial a},\frac{\partial S(w,t)}{\partial t},\frac{\partial D(w,a)}{\partial w},\frac{\partial S(w,t)}{\partial w}\bigg\},
\end{align*}
each of which is a real number representing a partial derivative describing the supply and demand function or a total derivative indicating a change over time within a scenario.   

An analysis of the two existential problems as suggested in Section \ref{SUBSEC:Framework}. shows that the result is \verb+Mixed+: that is both examples and counterexamples exist.  In particular, even when all of the assumptions are satisfied, it is possible that wages actually go down.  

Moreover, if $\frac{\partial D(w,a)}{\partial w}$ and $\frac{\partial S(w,t)}{\partial w}$ are left as free variables, a QE analysis recovers a disjunction of three quantifier-free formulae.  Two of them contradict the assumptions, but the third does not and can therefore be added to $A$ in order to guarantee the truth of $H$:
\[
\frac{\partial S(w,t)}{\partial w}\geq-\frac{\partial D(w,a)}{\partial w}>0.
\]
I.e. assuming that labour supply is at least as sensitive to wages as labour demand is (recall that the demand slope is a negative number) guarantees that wages $w$ are higher at the end of the recession that is primarily due to taxes on labour supply.   See also \cite{Mulligan2012} for further details on the economics.  

This example can be found in the benchmark set as Model \verb+#0013+.

\subsection{Vector summaries}
\label{SUBSEC:Hicks}

Economics problems sometimes involve an unspecified (and presumably large) number of variables.  Take Sir John Hicks' analysis of household decisions among a large number $N$ of goods and services \cite{Hicks1946}.  The quantities purchased are represented as a $N$-dimensional vector, as are the prices paid.  

We assume that when prices are $\bf p$ ($\hat{\bf p}$), the household makes purchases $\bf q$ ($\hat{\bf q}$), respectively:\footnote{The price change is compensated in the Hicksian sense, which means that $\bf q$ and $\hat{\bf q}$ are the same in terms of results or ``utility'', so that the consumer compares them only on the basis of what they cost (dot product with prices).}
\begin{equation*}
A \equiv (\bf p \cdot \bf q \le \bf p \cdot \hat{\bf q}) 
\, \wedge \,  
(\hat{\bf p} \cdot \hat{\bf q} \le \hat{\bf p} \cdot \bf q).
\end{equation*}
Hicks asserted that the quantity impact of the price changes $ \hat{\bf q} - \bf q$ cannot be positively correlated with the price changes $ \hat{\bf p} - \bf p$:
\begin{equation*}
	H \equiv (\hat{\bf q} - \bf q) \cdot (\hat{\bf p} - \bf p) \leq 0.
\end{equation*}
Hicks' reasoning depends only on the value of vector dot products, four of which appear above, rather than the actual vectors themselves whose length remains unspecified.
 
Hicks implicitly assumed that prices and quantities are real valued, which places additional restrictions on the dot products.  We need then to restrict that the symmetric Gram matrix corresponding to the four vectors $\hat{\bf q}, {\bf q}, \hat{\bf p}, {\bf p}$ be positive semi-definite.  To state this restriction we then need to increase the list of variables from the four dot products that appear in $A$ and $H$ to all ten that it is possible to produce with four vectors.  

QE technology certifies that there are no values for the dot products that can simultaneously satisfy $A$ and contradict $H$.  It follows that Sir John Hicks was correct regardless of the length $N$ of the price and quantity vectors.  In fact, further, it turns out that the additional restriction to real valued quantities is not needed for this example (there are no counter-examples even without this).  While those additional restrictions could hence be removed from the theorem we have left them in the statement in the benchmark set because (a) they would be standard restrictions an economist would work with and (b) it is not obvious that they are unnecessary before the analysis.

The above analysis could be performed easily with both \textsc{Mathematica} and \textsc{Z3} but proved difficult for \textsc{Redlog}.  We have yet to find a variable ordering that leads to a result in reasonable time\footnote{Since the problem is fully existentially quantified we are free to use any ordering of the 10 variables; but there are $10!=3,628,800$ such orderings; so we have not tried them all.}.  It is perhaps not surprising that a tool like \textsc{Z3} excels in comparison to computer algebra here, since the problem has a large part of the logical statement irrelevant to its truth: the search based SAT algorithms are explicitly designed to avoid considering this.

This example can be found in the benchmark set as Model \verb+#0078+.

%\newpage

\subsection{Concave production function}
\label{SUBSEC:JehleReny}

Our final example is adapted from the graduate-level microeconomics textbook \cite{JR11}.  The example asserts that any differentiable, quasi-concave, homogeneous, three-input production function with positive inputs and marginal products must also be a concave function.  In the Tarski framework, we have a twelve-variable sentence, which we have expressed simply with variable names $v_1, \dots, v_{12}$ to compress the presentation:

\begin{align*}
A &\equiv v_{1} v_{10}+v_{2} v_{7}+v_{3} v_{5}=0 \land v_{1} v_{11}+v_{2} v_{8}+v_{3} v_{7}=0
\\
&\quad 
\land v_{1} v_{12}+v_{10} v_{3}+v_{11} v_{2}=0 
\\
&\quad 
\land v_{1}>0 \land v_{2}>0 \land v_{3}>0 
\land v_{4}>0 \land v_{6}>0 \land	v_{9}>0 
\\
&\quad 
\land 2 v_{11} v_{6} v_{9}>v_{12} v_{6}^2+v_{8} v_{9}^2 
\\
&\quad 
\land 2 v_{10} v_{6} (v_{11} v_{4}+v_{7} v_{9})+v_{9} (2 v_{11} v_{4} v_{7}-2 v_{11} v_{5} v_{6}+v_{5} v_{8} v_{9})
\\
&\qquad 
	+ v_{12} \big(v_{4}^2 v_{8}-2 v_{4} v_{6} v_{7}+v_{5} v_{6}^2\big)>
%\\
%&\qquad \qquad 
	v_{10}^2 v_{6}^2+2 v_{10} v_{4} v_{8} v_{9}+v_{11}^2 v_{4}^2+v_{7}^2 v_{9}^2,
%\end{align*}
%\begin{align*}
\\
\quad
\\
H &\equiv 
v_{12}\leq 0\land v_{5} \leq 0\land v_{8}\leq 0 
\\
&\qquad \land
	v_{12} v_{5}\geq v_{10}^2 \land
	v_{12} v_{8}\geq v_{11}^2\land
	v_{8} v_{5}\geq v_{7}^2
\\
&\qquad \land 
	v_{8} \left(v_{10}^2-v_{12} v_{5} \right)+v_{11}^2 v_{5}+v_{12} v_{7}^2\geq 2 v_{10} v_{11} v_{7} .
\end{align*}

In disjunctive normal form (DNF) $A \wedge \neg H $ is a disjunction of seven clauses.  Each atom of $H$, and therefore each clause of $A \wedge \neg H $ in DNF, corresponds to a principal minor of the production function's Hessian matrix.  So the clauses in the DNF are $A \land v_{12}>0, A \land v_5 >0 \land \dots$ in the same sequence used in $H$ above.  The existential quantifiers applied to $A \wedge \neg H $ can be distributed across the clauses to form seven smaller QE problems.

As of early 2018, \textsc{Z3} could not determine whether the Tarski formula $A \wedge \neg H $ is satisfiable: it was left running for several days without producing a result or error message.  For this problem virtual substitution techniques may be advantageous (note the degree in any one variable is at most two).  Indeed, \textsc{Redlog} and \textsc{Mathematica} can resolve it quickly.  The problem would certainly be difficult for CAD based tools (which underline the theory solver used by \textsc{Z3} for such problems).  For example, if we take only the first clause of the disjunction (the one containing $v_{12} > 0$), then the Tarski formula has twelve polynomials in twelve variables, and just three CAD projections (using Brown's projection operator) to eliminate $\{v_{12} , v_{11}, v_{10}\}$ results in 200 unique polynomials with nine variables still remaining to eliminate!

We note that the problem has a lot of structure to exploit in regards to the sparse distribution of variables in atoms.  One approach currently under investigation is to eliminate quantifiers incrementally, taking advantage of the repetitive structure of the sub-problems one uncovers this way (an approach that works on several other problems in the benchmark set).  This will be the topic of a future paper.

This example can be found in the benchmark set as Model \verb+#0056+.

\section{Analysis of the Structure of Economics Benchmarks}
\label{SEC:ExampleCollection}

The examples collected for the benchmark set come from a variety of areas of economics.  They were chosen for their importance in economic reasoning and their ability to be expressed in the Tarski framework, not on the basis of their computational complexity.  The sentences tend to be significantly larger formulae than the initial examples described above.  We summarise some statistics on the benchmarks.  These statistics are generated for the 45 counterexample checks ($\exists \V[A \land \lnot H]$) which are representative of the full 135 problems.

\subsection{Logical size}

Presented in DNF the benchmarks have number of clauses ranging from 7 to 43 with a mean average of 18.5 and median of 18.  Of course each clause can be a conjunction but this is fairly rare:  the average number of literals in a clause is at most 1.5 in the dataset, with the average number of atoms in a formula only slightly more than the number of clauses at 19.3.  

\subsection{Algebraic size}

The number of polynomials involved in the benchmarks varies from 7 to 43 with both mean average of 19.2 and median average of 19.  The number of variables ranges from 8 to 101 with an mean average of 17.2 and median of 14.  It is worth noting that examples with this number of variables would rarely appear in the QE literature, since many QE algorithms have complexity doubly exponential in the number of variables \cite{EBD15}, \cite{ED16a}, with QE itself doubly exponential in the number of quantifier changes \cite{BPR06}.  The number of polynomials per variable ranges from 0.4 to 2.0 with an average of 1.2.

\subsection{Low degrees}

While the algebraic dimensions of these examples are large in comparison to the QE literature, their algebraic structure is usually less complex.  In particular the degrees are strictly limited.
The maximum total degree of a polynomial in a benchmark  ranges from 2 to 7 with an average of 4.2.  So although linear methods cannot help with any of these benchmarks the degrees do not rise far.  The mean total degree of a polynomial in a benchmark is only 1.8 so the non-linearity while always present can be quite sparse.

However, of greater relevance to QE techniques is not total degree of a polynomial but the maximum degree in any one variable: e.g. this controls the number of potential roots of a polynomial and thus cell decompositions in a CAD (with this degree becoming the base for the double exponent in the complexity bound); see for example the complexity analysis in \cite{BDEMW16}.  The maximum degree in any one variable is never more than 4 in any polynomial in the benchmarks.  If we take the maximum degree in any one variable in a formula and average for the dataset we get 2.1; if we take the maximum degree in any one variable in a polynomial and average for the dataset we get 1.3.

\subsection{Plentiful useful side  conditions}

With variables frequently multiplying each other in a sentence's polynomials, a potentially large number of polynomial singularities might be relevant to a QE procedure.  However, we note that the benchmarks here will typically include sign conditions that in principle rule out the need to consider many such singularities.  That is, these side conditions will often coincide with unrealistic economic interpretations that are excluded in the assumptions.  Of course, the computational path of many algorithms may not necessarily exploit these currently without human guidance.

\subsection{Sparse occurrence of variables}

For each sentence $\exists \V[A \land \lnot H]$ (these are the computationally more challenging of the 135) we have formed a matrix of occurrences, with rows representing variables and columns representing polynomials.  A matrix entry is one if and only if the variable appears in the polynomial, and zero otherwise.  The average entry of all 45 occurrence matrices is 0.15.  
This sparsity indicates that the use of specialised methods or optimisations for sparse systems could be beneficial.

\subsection{Variable orderings}

Since the problems are fully existentially quantified we have a free choice of variable ordering for the theory tools.  It is well known that the ordering can greatly affect the performance of tools.  A classic result in CAD presents a class of problems in which one variable ordering gave output of double exponential complexity in the number of variables and another output of a constant size \cite{BD07}.  Heuristics have been developed to help with this choice (e.g. \cite{DSS04} \cite{BDEW13}) with a machine learned choice performing the best in \cite{HEWDPB14}.  The benchmarks come with a suggested variable ordering but not a guarantee that this is optimal: tools should investigate further (see also \cite{MDE18}).

\section{Final Thoughts}
\label{SEC:Final}

\subsection{Summary}

We have presented a dataset of examples from economic reasoning suitable for automatic solution by QE or SMT technology.  These clearly broaden the application domains present in the relevant category of the SMT-LIB, and the literature of QE.  Further, their logical and algebraic structure are sources of additional diversity.  

While some examples can be solved easily there are others that prove challenging to the various tools on offer.  The example in Section \ref{SUBSEC:Hicks} could not be tackled by \textsc{Redlog} while the one in Section \ref{SUBSEC:JehleReny} caused problems for \textsc{Z3}. Of the tools we tried, only \textsc{Mathematica} was able to decide all problems in the benchmark set.  
%In fact all 135 could be tackled by \textsc{Mathematica} in less than a minute on a laptop computer, with only three of those taking more than ten seconds, pointing to a string suite of techniques underlying the \texttt{Resolve} command which makes a good choice of which to use.
However, in terms of median average timing over the benchmark set \textsc{Mathematica} takes longer with $825$ms than \textsc{Redlog} ($50$ms) or \textsc{Z3} ($<15$ms).  

\subsection{Ongoing and future work}

As noted at the end of Section \ref{SEC:Examples}, some of these problems have a lot of structure to exploit and so the optimisation of existing QE tools for problems from economics is an ongoing area of work.  

More broadly, we hope these benchmarks and examples will promote interest from the Computer Science community in problems from the social sciences, and correspondingly, that successful automated solution of these problems will promote greater interest in the social sciences for such technology.  A key barrier to the latter is the learning curve that technology like Computer Algebra Systems and SMT solvers has, particularly to users from outside Mathematics and Computer Science.  In \cite{MDE18} we describe one solution for this: a new \textsc{Mathematica} package called \texttt{TheoryGuru} which acts as a wrapper for the QE functionality in the \texttt{Resolve} command designed to ease its use for an economist.  
%It can parse input perform a number of error checks, check assumptions and add standard ones that may have been forgotten, make choices about how to run the algorithm (e.g. variable ordering), and offer interpretations on the output.  

\bibliographystyle{plain}
\bibliography{CAD}

%------------------------------------------------------------------------------

\end{document}